\documentclass[12pt]{article}
\usepackage{a4wide,amssymb}
\usepackage{amsmath}
\usepackage{cite}
\usepackage{latexsym}
\usepackage{graphicx}
\usepackage{psfrag}

\def\Ai{{\rm Ai}}

\def\1{\bf{1}}
\def\R{\mathbb{R}}

\def\k{\kappa}

\begin{document}

\title{Exact solution for the stationary \\ Kardar-Parisi-Zhang equation}
\author{Takashi Imamura
\footnote {Research Center for Advanced Science and Technology,
The University of Tokyo, 4-6-1 Komaba, Meguro, Tokyo 153-8904, Japan,
imamura@jamology.rcast.u-tokyo.ac.jp},
Tomohiro Sasamoto
\footnote{Department of Mathematics and Informatics, 
 Chiba University, 1-33 Yayoi-cho, Inage, Chiba 263-8522, Japan,
sasamoto@math.s.chiba-u.ac.jp}}
\maketitle

\begin{abstract}
We obtain the first exact solution for the stationary one-dimensional 
Kardar-Parisi-Zhang equation. A formula for the distribution of the height is given in terms 
of a Fredholm determinant,  which is valid for any finite time $t$. The 
expression is explicit and compact enough so that it can be evaluated 
numerically. 
Furthermore, by extending the same scheme, we find an exact 
formula for the stationary two-point correlation function. 
\end{abstract}

The dynamics of a growing surface has long been a subject of intense study 
because of its wide appearance in nature and relevance to material 
science; it is also of great theoretical interest as an important example 
of nonlinear nonequilibrium phenomena \cite{BS1995,Meakin1998}. 
For a surface growing only through local interaction, a prototypical 
equation was proposed by Kardar, Parisi and Zhang (KPZ) in 1986 ~\cite{KPZ1986}.
In this Letter, we focus on its one-dimensional version,
\begin{equation}
\frac{\partial h(x,t)}{\partial t}=\frac{\lambda}{2}
\left(\frac{\partial h(x,t)}{\partial x}\right)^2+
\nu\frac{\partial^2 h(x,t)}{\partial x^2}+\eta(x,t).
\label{KPZ}
\end{equation}
Here $h(x,t)$ represents the height of the surface at position 
$x\in \R$ and time $t\ge 0$.
The last term $\eta (x,t)$ is taken to be the Gaussian white noise 
with covariance
$\langle \eta(x,t)\eta(x',t')\rangle=D \delta(x-x')\delta(t-t')$.

This KPZ equation has been well known and has already been studied 
extensively~\cite{FNS1977,BG1997,CM2001,KS2004,W2009}, but interest in this equation has revived~\cite{TSSS2011}.
First, the efforts to have a better understanding of the equation,
which are from various perspectives and have lasted for more than two decades,
finally culminated in 
the achievement of an exact solution for its height distribution
\cite{SS2010,ACQ2010}; second it has turned out that such fine information as the 
height distribution can be measured experimentally. In 
\cite{TS2010}, Takeuchi and Sano employed a turbulent liquid 
crystal as a material and found a clear agreement with theoretical predictions. 

In the studies of growing surface, the fluctuations of the height are of 
primary importance. In many examples one observes the ``kinetic roughening."
As time goes on the surface becomes rougher and rougher,  
even if it starts from a completely flat substrate.
This concerns the transient behaviors which are certainly 
interesting, but even more natural and important is the stationary situation. 
Most systems eventually reach their own steady states 
after relaxation times and 
our understanding of the nonequilibrium steady state (NESS)  
is in general so limited that concrete results for a simple 
system should give us valuable information. 
Many papers on the KPZ equation have addressed its stationary properties 
with various methods but all attempts to treat this case exactly have so far failed. 

In this article, 
we provide the first exact results for the KPZ equation in the stationary situation. 
For the KPZ equation,
the stationary state is given by the two-sided Brownian motion (BM) 
with respect to $x$
\cite{KS1992}: $h(x,0) = 
   \alpha B_-(-x) {\text{~for~}} x<0,
   \alpha B_+(x){\text{~for~}} x>0,
$
where $\alpha=(2\nu)^{-3/2}\lambda D^{1/2}$ and $B_{\pm}(x)$ are two independent standard BMs, as seen in Fig.~1.
Since the system is translationally invariant, one can set 
without loss of generality the initial height at the origin to be zero, 
i.e., $h(0,0)=0$. In the following, we will show the explicit formulas for the height distribution 
${\rm Prob}[h(x,t)\geq s]$ and the two-point space-time correlation functions
of the slope $\langle \partial_x h(x,t) \partial_x h(0,0)\rangle$ (Fig.~1).  
The obtained 
expressions are explicit enough so that they can be evaluated numerically.
We would say that the main achievements in our paper are that we could draw 
their figures as given in Figs. 2 and 3 below.


\begin{figure}[h]
\begin{center}
 \includegraphics[scale=.5]{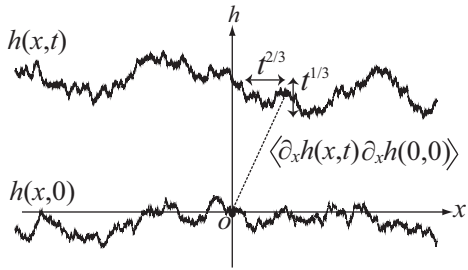}
\caption{The two-sided BM initial condition and the two-point function}
\end{center}
\end{figure}


Let us briefly recall the theoretical developments so far. 
Early works using renormalization group ideas and other analytic methods have 
identified the critical exponents. It was established that when $t$ is large, 
the fluctuations of the height grow like $O(t^{1/3})$ and the nontrivial 
equal-time correlations in space are observed on the $O(t^{2/3})$ scale. 
About a decade ago, it was realized that a few growth models admit a 
mapping to a combinatorial problem, which allowed us to study their height distributions and their universal scaling limits.
 In these models some of the space, time, and height are 
discrete; there is a natural ultraviolet cutoff, which makes their analysis  somewhat easier. The most surprising at the time was that the universal 
height distribution is the same as that of the largest eigenvalue of 
Gaussian ensembles from random matrix theory\cite{PS2000a,Jo2000}. 

But it was only very recently that these developments finally led to the 
identification of the exact height distributions for the KPZ equation itself
\cite{SS2010,ACQ2010}. 
It contains valuable information about finite time behaviors, which has significant relevance to real physical
systems because a slow approach to the 
limiting values for the moments has been 
observed in experiments as well. 
For more recent developments, see, for instance, \cite{CCDW2011p,
COSZ2011p,BC2011p}.

A most important finding from the exact solution is that the KPZ equation
represents the universal crossover between the linear ($\lambda=0$) Edwards-Wilkinson 
(EW) \cite{EW1982} and the nonlinear KPZ universality classes.
This is realized, for instance, in the fact that the KPZ equation is obtained 
as a weakly asymmetric limit of the asymmetric simple exclusion process (ASEP).
In other words, the KPZ equation describes the behaviors of the ASEP
in a certain time regime~\cite{BG1997}. 
The ASEP is a stochastic system of many particles, 
each of which performs an asymmetric random walk under the constraint 
of volume exclusion with other particles and is one of the most well-studied 
nonequilibrium discrete models. 
The KPZ equation also captures appropriately low temperature behaviors of  
some discrete random directed polymer models in ~\cite{CLDR2010,AKQ2010}, 
demonstrating the universality of the equation.



The first exact solution of the KPZ equation was obtained for the
narrow wedge initial condition, for which the shape of the surface 
at finite time is a parabola \cite{SS2010,ACQ2010}.
On the other side from the previous studies, 
it has been well established that the height 
distribution depends strongly on the geometry of the system and its 
initial conditions \cite{BR2001c,PS2000a}. 
In particular it is crucial whether the macroscopic shape of the surface is curved or flat. 
This was confirmed in an experiment as well \cite{TSSS2011}. 
Hence, it is obviously important to extend the exact results to other cases,
which is not at all trivial. 
The analysis in 
\cite{SS2010,ACQ2010} 
was based on a previous exact result of the ASEP~\cite{TW2009a}.
The ASEP has been extensively 
studied, but the studies of its fluctuation properties are fairly 
involved, and there are not so many generalizations obtained so far.

In the meantime the same expression for the narrow wedge initial condition 
was rederived by using a replica method 
\cite{CLDR2010,Dotsenko2010}. The
replica method is well known to be powerful for studying systems with 
randomness but is also known to possess a serious difficulty related 
to the analytic continuation for the replica number. 
Interestingly enough, for the KPZ equation, though there appear divergent sums during 
the computations one finally arrives at the correct finite expression. Hence, 
a replica analysis of the KPZ problem should be useful for clearer 
understanding of replica method in general. Moreover, this approach has turned out to be 
more suited for various generalizations. In fact multipoint 
distributions were studied in \cite{PS2011} 
and, more recently, the flat initial condition was treated in \cite{CLD2011}. 
In the present Letter, we study the stationary case by a nontrivial 
extension of this replica approach.  
By a set of scalings of space, time, and height, 
$ x \to \alpha^2 x, \quad t \to 2\nu \alpha^4 t, 
 \quad h \to \frac{\lambda}{2\nu}h$, 
we can and will do set 
$ \nu = \frac12, -\lambda = D=1$.

Now, we start to explain our solutions. The main object in this Letter is
the height distribution function defined by 
\begin{equation}
F_{v_{\pm},t}(s)=\text{Prob}\left(h(0,t)-t/24\ge \gamma_t s\right)
\end{equation}
where $\gamma_t=(t/2)^{1/3}$.
Note that this reflects previous studies: The average is given by $\langle h(x,t) \rangle = t/24$, and the fluctuation scales like $O(t^{1/3})$ 
from the KPZ scaling. 
It has not been clear how one can take the average over the BM initial 
condition 
directly, or if it is possible at all. The novel strategy here in studying 
the stationary state 
is that we first consider a generalized initial condition and  then retrieve the desired information from the results for it. The initial condition we consider is 
\begin{equation}
 h(x,0) = \begin{cases}
	   B_-(-x) - v_- x, & x<0,\\
           B_+(x) + v_+ x,  & x>0,
	  \end{cases}
\label{init_drift}
\end{equation}
where 
$v_{\pm}$ are the strength of the drifts. Once this generalized 
case is solved, then the stationary case can be accessed by 
taking the $v_{\pm}\to 0$ limit. 

Applying the Cole-Hopf transformation, 
$ Z(x,t) = e^{-h(x,t)}$,
the KPZ equation (\ref{KPZ}) is linearized as 
\begin{equation}
\frac{\partial Z(x,t)}{\partial t}=
\frac12 \frac{\partial^2 Z(x,t)}
{\partial x^2}-\eta(x,t)Z(x,t).
\label{SHE}
\end{equation}
This can be interpreted as a problem of a directed polymer in 
random media; $Z$ is its partition function. 
We consider the $N$th replica partition function 
$\langle Z^N(x,t)\rangle$ and compute their generating 
function $G_t(s)$ defined as
\begin{equation}
G_t(s)
=
\sum_{N=0}^{\infty}\frac{\left(-e^{-\gamma_t s}\right)^N}{N!}
\left\langle 
Z^N (0,t)
\right\rangle  
e^{N\frac{\gamma_t^3}{12}}.
\label{gr}
\end{equation}

Using the Feynman path integral representation of $Z$
and taking the average with respect to the random potential $\eta$, 
the replica partition function can be written in a form, 
$\langle Z^N(x,t) \rangle=\langle x|e^{-H_Nt}|\Phi\rangle$.
Here 
$\langle x|$ represents the state with all $N$ particles 
being at the position $x$ and $|\Phi\rangle$, the initial state.  
For the KPZ equation, the Hamiltonian $H_N$ turns out to be that of 
the delta-function Bose gas with attractive interaction 
\cite{Kardar1987}, 
\begin{equation}
H_N=-\frac{1}{2}\sum_{j=1}^N\frac{\partial^2}{\partial x_j^2}
-\frac12\sum_{j\neq k}^N\delta(x_j-x_k).
\label{dBose}
\end{equation}
The eigenvalues and eigenfunctions
can be constructed by using the Bethe 
ansatz~\cite{LL1963,McG1964,Dotsenko2010,CLDR2010}. 
For a set of quasimomenta $z_j$'s, the eigenvalue $E_z$ is given by
$E_z=\frac12\sum_{j=1}^N z_j^2$ 
and the eigenfunction by 
$\langle x_1,\cdots, x_N|\Psi_z\rangle
=C_z\sum_{P\in S_N}{\text{sgn}P}\prod_{1\le j<k\le N}
\left(z_{P(j)}-z_{P(k)}\right.$ \\
$\left.+i\text{sgn}(x_j-x_k)\right)\exp\left(i\sum_{l=1}^N
z_{P(l)}x_l\right)$,
where $S_N$ is the set of permutations with $N$ elements and
$C_z$ is the normalization constant.
For the $\delta$-Bose gas with attractive interaction,
the quasimomenta $z_j~(1\le j\le N)$ are in general complex numbers
and are aligned in the form of ``strings"~\cite{Dotsenko2010}.
The completeness of these Bethe states was proved very recently 
\cite{PS2011p2}. 
We will compute this moment by expanding as 
$\langle Z^N(x,t) \rangle=\sum_{z} e^{-E_z t}\langle x|\Psi_z\rangle \langle \Psi_z|\Phi\rangle$.

Because the Brownian motion is a Gaussian process, 
one can perform the average over the initial distribution 
(\ref{init_drift})
and the inner product of $\langle\Psi_z|\Phi\rangle$ can 
also be explicitly calculated under the condition $v_{\pm}>0$ .
It includes the summation over $S_N$ coming from the Bethe eigenstate, 
which looks highly nontrivial but  
we have found that there is a combinatorial formula which factorizes it~\cite{IS2011p}.
We obtain
\begin{align}
&~\langle \Psi_z|\Phi\rangle
=N!C_z \frac{ \prod_{m=1}^N (v_++v_-- m)  \prod_{1\le j<k\le N}(z^*_j-z^*_k)}
   { \prod_{m=1}^N(-iz^*_m+v_--1/2) (-iz^*_m-v_++1/2)}.
\label{nodangerous}
\end{align}
Now we can follow the arguments in \cite{IS2011}.  
We further modify the expression of the replica partition function 
and write it in terms of a determinant by using the Cauchy's 
determinant formula.
After some computation,  
one arrives at an expression for the generating function.

For the narrow wedge \cite{SS2010,ACQ2010} and the half BM initial 
condition \cite{IS2011}, this generating 
function itself is written as a Fredholm determinant, but this time it 
is not because of the novel factor $\prod_{l=1}^N(v_++v_--l)$. 
The difficulty can be avoided by considering a further generalized 
initial condition for the KPZ equation in which the initial 
overall height $\chi$ obeys a certain probability distribution. 
A similar argument was already used for other  discrete models 
\cite{BR2001c,IS2004,FS2006}.
If we denote the height with this initial distribution by $\tilde{h}$,  
we have 
 $\tilde{h} = h - \chi$, 
where $h$ is the original height for which $h(0,0)=0$. 
The random variable $\chi$ is taken to be independent of $h$ and hence  
the $N$th moment of $e^{-\tilde{h}}$ factorizes as 
$ \langle e^{-N\tilde{h}} \rangle
 =
 \langle e^{-N h} \rangle \langle e^{N\chi} \rangle$.
Moreover, the probability distribution functions  
$\tilde{F}(s)=\text{Prob}[\tilde{h}(0,t)\geq \gamma_t s]$ and  
$F(s)=\text{Prob}[h(0,t)\geq \gamma_t s]$ 
are related as 
$ F(s)=\frac{1}{ \kappa(\gamma_t^{-1}\frac{d}{ds})}\tilde{F}(s),$
where $\k$ is the Laplace transform of the pdf of $\chi$, and 
we assume $1/\kappa(\xi)$ can be Taylor expanded around $\xi=0$.  
Here it should be noticed that if $e^{\chi}$ 
obeys the inverse gamma distribution 
with parameter $v_++v_-$, i.e., if $e^{-\chi}$ 
obeys the gamma distribution with 
the same parameter, its $N$th moment is given by 
$1/\prod_{l=1}^N(v_++v_--l)$, which exactly compensates for the extra factor above. 
Hence the generating function $\tilde{G}_t(s)$ 
corresponding to $\tilde{h}$ can be written as a Fredholm determinant.

Following the arguments in \cite{CLDR2010,PS2011,IS2011}, 
one can recover the information of the distribution for $\tilde{h}$
from the generating function $\tilde{G}_t$.
Combining this and a fact
 $\kappa(\xi) = \Gamma(v+\xi)/\Gamma(v)$
for the case where $e^{\chi}$ is the 
inverse gamma random variable with parameter $v$,
we find that the height distribution for the initial condition (\ref{init_drift}) 
is given by  
\begin{align}
F_{v_{\pm},t}(s)&
=
\frac{\Gamma(v_++v_-)}{\Gamma(v_++v_-+\gamma_t^{-1}d/ds)}
\left[
1-\int_{-\infty}^{\infty}
du e^{-e^{\gamma_t(s-u)}}
\nu_{v_{\pm},t}(u)
\right] . 
\label{Result}
\end{align}
Here $\nu_{v_{\pm},t}(u)$ is expressed as a difference of two
Fredholm determinants, 
\begin{align}
\nu_{v_{\pm},t}(u)
=\det\left(1-P_u(B^{\Gamma}_{t}-P^{\Gamma}_{\Ai})P_u\right)
 -\det\left(1-P_uB^\Gamma_{t}P_u\right),
\end{align}
where $P_s$ represents the projection onto $(s,\infty)$,
\begin{align}
P^{\Gamma}_{\Ai}(\xi_1,\xi_2)
&=
\Ai_\Gamma^\Gamma\left(\xi_1,
\frac{1}{\gamma_t},v_-,v_+\right)
\Ai_\Gamma^\Gamma \left(\xi_2,
\frac{1}{\gamma_t},v_+,v_-\right), \\
B^{\Gamma}_{t}(\xi_1,\xi_2)
&=
\int_{-\infty}^{\infty} dy \frac{1}{1-e^{-\gamma_t y}} 
\Ai_\Gamma^\Gamma\left(\xi_1+y,
\frac{1}{\gamma_t},v_-,v_+\right)
\Ai_\Gamma^\Gamma \left(\xi_2+y,
\frac{1}{\gamma_t},v_+,v_-\right),
\label{BGamma}
\end{align}
and
\begin{align}
&\Ai_{\Gamma}^\Gamma (a,b,c,d)
=\frac{1}{2\pi}
\int_{\Gamma_{i\frac{d}{b}}} dz 
e^{iza+i\frac{z^3}{3}}
\frac{\Gamma\left(ibz+d\right)}{\Gamma\left(-ibz+c\right)},
\label{AiGG}
\end{align}
where
$\Gamma_{z_p}$ represents the contour  from 
$-\infty$ to $\infty$ which, along the way, 
passes below the pole at $z=id/b$.


With this expression it is not hard to do analytic continuation for $v_{\pm}$ to their 
negative regions and then take the $v_{\pm}\to 0$ limit to study the stationary 
situation. Our result for the height distribution for the stationary KPZ equation is  
written as 
\begin{equation}
{\rm Prob}\left[h(0,t)-t/24\geq \gamma_t s\right]
=
\frac{1}{\Gamma(1+\gamma_t^{-1}d/ds)}
\int_{-\infty}^{\infty} du \gamma_t  e^{\gamma_t(s-u) + e^{-\gamma_t(s-u)}}\nu_{0,t}(u).
\label{Result}
\end{equation}
Here $\nu_{0,t}(u)$ is obtained from $\nu_{v_{\pm},t}(u)$ by taking $v_{\pm}\to 0$ limit
as in \cite{FS2006}. Details will be given elsewhere \cite{IS2011p}. 
By using this formula, one can numerically evaluate the distribution. 
In Fig. 2, we plot this distribution (\ref{Result}) for $\gamma_t=1$.  
Once we have the formula for finite $t$, it is not difficult to consider the 
scaling limit. In our setting, it can be studied simply by taking $t\to\infty$.
One can confirm that it tends to $F_0$ as given in \cite{FS2006}.
This function already appeared in the studies of lattice growth models
and describes the stationary height distribution in the scaling limit
\cite{PS2002a,PS2004,FS2006}. 

%


\begin{figure}[h]
\begin{center}
 \includegraphics[scale=.6]{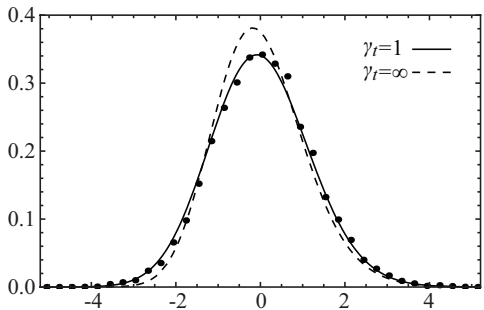}
\caption{Stationary height distribution for the KPZ equation for 
$\gamma_t=1$. The dashed curve is $F_0$. The dots ($\cdot$) indicate
the Monte Carlo simulation for WASEP.
We set the difference between left and right hopping rates 0.15 
with their sum normalized to be unity,
$3951$ Monte Carlo steps (corresponding to $\gamma_t=1$), and 10000 samples.
}
\end{center}
\end{figure}


It is not hard to generalize our analysis to the height near the 
origin. From the KPZ scaling, the nontrivial correlation is expected 
on $O(t^{2/3})$ scale, and we can see that the height distribution 
at $x=2\gamma_t^2X $ with $X$ finite is written as the same form as in 
(\ref{Result}) with its kernel slightly modified to include $X$. 
This can be readily used to obtain the exact formula for the stationary 
two-point correlations. Let us set   
$C(x,t) = \langle (h(x,t)-\langle h(x,t)\rangle)^2 \rangle$. 
One can check  that the second derivative of this function with respect to 
$x$  is, in fact, twice the two-point correlation: 
$\partial_x^2 C(x,t) = 2 \langle \partial_x h(x,t) \partial_x h(0,0)\rangle$. 
From the KPZ scaling, we also introduce the scaled version, 
$g_t(y) = (2t)^{-2/3} C \left( (2t)^{2/3}y,t \right)$. 
Plotted in Fig. 3 is the second derivative of this function $g_t(y)$.  
In the scaling limit, it tends to the function $g(s)$,
which was studied before in \cite{PS2002a,PS2004,FS2006,BFP2010}. 
More discussions will be given in \cite{IS2011p}. 

Here we remark that 
in statistical mechanics, computing the dynamical correlation functions
is one of the most important objectives of the theory. For instance, in 
the Kubo formula, one needs 
the dynamical current-current correlation function in equilibrium. 
This is already known to be very difficult in general and it is even 
more interesting and challenging to compute the dynamical correlation 
for nonequilibrium systems, in particular in its stationary state. In addition,
the Fourier transform of the two-point function is known as the structure 
function, which can be measured directly using (e.g., neutron) scattering 
experiments. 

\begin{figure}[h]
\begin{center}
\includegraphics[scale=.6]{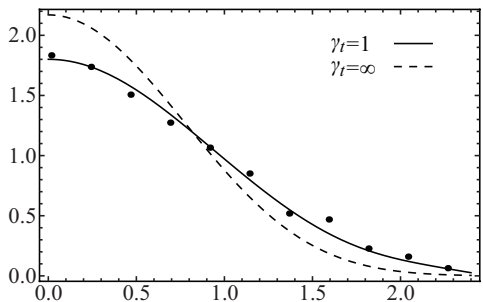}
\caption{Stationary two-point correlation function $g_t''(y)$ for the 
KPZ equation for $\gamma_t=1$. The dashed curve represents the one 
in the scaling limit $g''(y)$. The dots indicates the Monte Carlo result of the WASEP, where parameters are the same as those in Fig. 2.}
\end{center}
\end{figure}

Our exact results represent the universal crossover 
for the KPZ equation
with $t$ being a parameter. As explained above, in a certain time regime,
the fluctuation properties of the weakly ASEP (WASEP) 
are described by the KPZ equation.
We confirmed this by comparing our theoretical predictions
with Monte Carlo simulations of the WASEP (Figs. 2 and 3). The dots in
Figs. 2 and 3 represent the Monte Carlo results for 
the current distribution and density-density  
correlation in the WASEP respectively. 

To summarize, we have found the first explicit formulas for the height 
distribution and the two-point correlation functions for the KPZ equation 
in its stationary regime.  They are explicit enough to allow numerical 
evaluations as depicted in Figs. 2 and 3. One observes nontrivial 
deviations from the scaling limit. They contain valuable information 
about the finite $t$ corrections which should be observed in real 
experiments. One possibility to realize the stationary situation in an 
experiment is to use again the liquid crystal turbulence because the 
system is fairy flexible for setting a desired initial condition by adjusting a shape 
of the laser pulse. It might also be possible to observe our prediction 
by performing a careful experiment of paper combustion for which 
the stationary situation was already studied in a previous publication
\cite{Miettinen2005}. 

\section*{Acknowledgments}
T.S. thanks  
I. Corwin, P.L. Ferrari, S. Prolhac, J. Quastel, and H. Spohn  for useful discussions.
Both authors would like to thank R.Y. Inoue 
for enjoyable conversations on related issues. 
The work of T.I. and T.S. is supported by 
KAKENHI (22740251) and KAKENHI (22740054), respectively.


\begin{thebibliography}{10}

\bibitem{BS1995}
A.L. Barab{\'a}si and H.E. Stanley, 
\newblock {\em Fractal concepts in surface growth}
\newblock (Cambridge University Press, 1995).

\bibitem{Meakin1998}
P.~Meakin.
\newblock {\em Fractals, scaling and growth far from equilibrium}
\newblock (Cambridge University Press, 1998).

\bibitem{KPZ1986}
{M. Kardar, G. Parisi, and Y. C. Zhang},
\newblock {\em Phys. Rev. Lett.}  {\bf 56}, 889
(1986).

\bibitem{FNS1977}
{D.~Forster, D.~R.~Nelson, and M.~J.~Stephen},
\newblock{\em Phys. Rev. A.} {\bf 16}, 732
(1977).

\bibitem{BG1997}
L.~Bertini and G.~Giacomin,
\newblock{Commun. Math. Phys.}, {\bf 183}, 571
(1997).

\bibitem{CM2001}
{F.~Colaiori and M.~A.~Moore},
\newblock{\em Phys. Rev. Lett.} {\bf 86}, 3946
(2001).

\bibitem{KS2004}
{E.~Katzav and M.~Schwartz},
\newblock{\em Phys. Rev. E} {\bf 69}, 052603
(2004).

\bibitem{W2009}
{H.~S.~Wio},
\newblock{\em Int. J. Bifurcation Chaos Appl. Sci.
Eng.}{\bf 19}, 2813
(2009).

\bibitem{TSSS2011}
{K. A. Takeuchi, M. Sano, T. Sasamoto, and H. Spohn},
\newblock {\em Sci. Rep.} {\bf 1}, 34 (2011).

\bibitem{SS2010}
T.~Sasamoto and H.~Spohn,
\newblock {\em Phys. Rev. Lett.} {\bf 104}, 230602 
(2010); 
\newblock {\em J. Stat. Phys.}  {\bf 140}, 209 
(2010);
\newblock {\em Nucl. Phys. B}  {\bf 834}, 523 
(2010).



\bibitem{ACQ2010}
{G. Amir, I. Corwin and J. Quastel},
\newblock{\em Commun. Pure Appl. Math.}  {\bf 64}, 466 
(2011).

\bibitem{TS2010}
{K.A. Takeuchi and M. Sano},
 \newblock{\em Phys. Rev. Lett.} {\bf 104}, 230601 (2010). 

\bibitem{KS1992}
J.~Krug and H.~Spohn,
\newblock {\em Solids far from Equilibrium: Growth,
  Morphology and Defects} edited by C.~Godr\`eche 
(Cambridge University Press, Cambridge, 1992),p. 
479. 

\bibitem{Jo2000}
K.~Johansson,
\newblock{\em Commun. Math. Phys.} {\bf 209}, 437 (2000).

\bibitem{PS2000a}
M.~Pr{\"a}hofer and H.~Spohn,
\newblock {\em Phys. Rev. Lett.}  {\bf 84},  4882 
(2000).

\bibitem{CCDW2011p}
{L. Canet, H. Chat{\'e}, B. Delamotte and N. Wschebor},
\newblock{\em Phys. Rev. Lett.}  {\bf 104}, 150601 
(2010); Phys. Rev. E {\bf 84}, 061128 (2011).
arXiv:1107.2289.


\bibitem{COSZ2011p}
{I. Corwin, N. O'Connell, T. Sepp\"al\"ainen, and N. Zygouras},
arXiv:1110.3489.

\bibitem{BC2011p}
{A. Borodin and I. Corwin},
arXiv:1111.4408.

\bibitem{EW1982}
{S.F. Edwards and D.R. Wilkinson},
\newblock{\em Proc. R. Soc. A}  {\bf 381}, 17 
(1982). 

\bibitem{CLDR2010}
{ P. Calabrese, P. Le Doussal and A. Rosso},
\newblock {\em Euro Phys. Lett.}  {\bf 90}, 20002 (2010).

\bibitem{AKQ2010}
T.~Alberts, K.~Khanin, and J.~Quastel,
\newblock {\em Phys. Rev. Lett.}  {\bf 105}, 090603 (2010).

\bibitem{BR2001c}
J.~Baik and E.~M. Rains,
\newblock in {\em Random Matrix Models and Their Applications}, 
edited by P.~M. Bleher and A.~R. Its (Cambridge University Press, Cambridge,
2001).


\bibitem{TW2009a}
C.~A. Tracy and H.~Widom,
\newblock {\em Commun. Math. Phys.}  {\bf 209}, 129 
(2009).




\bibitem{Dotsenko2010}
{ V. Dotsenko},
\newblock {\em Euro Phys. Lett.}  {\bf 90}, 200003 (2010).

\bibitem{PS2011}
S.~Prolhac and H.~Spohn,
\newblock {\em J. Stat. Mech.}  P01031 (2011);
\newblock{\em J. Stat. Mech.} P03020 (2011).


\bibitem{CLD2011}
{ P. Calabrese and P. Le Doussal},
\newblock {\em Phys. Rev. Lett.}  {\bf 106} 250603  (2011).

\bibitem{Kardar1987}
M.~Kardar,
\newblock {\em Nucl. Phys. B.}  {\bf 290}, 582 
(1987).

\bibitem{LL1963}
{E. H. Lieb and W. Liniger},
\newblock {\em Phys. Rev.}  {\bf 130}, 1605 
(1963).

\bibitem{McG1964}
J.~B. McGuire,
\newblock {\em J. Math. Phys.} {\bf 5}, 622 
(1964).



\bibitem{PS2011p2}
S.~Prolhac and H.~Spohn,
\newblock {\em J. Math. Phys.} {\bf 52}, 122106 (2011).


\bibitem{IS2011}
T.~Imamura and T.~Sasamoto,
\newblock{\em J. Phys. A: Math. Theor.}  {\bf 44} 385001 (2011). 

\bibitem{IS2004}
T.~Imamura and T.~Sasamoto, 
\newblock {\em Nucl. Phys. B}  {\bf 699}, 503 
(2004).

\bibitem{IS2011p}
T.~Imamura and T.~Sasamoto, (in preparation). 

\bibitem{FS2006}
{P. L. Ferrari and H. Spohn},
\newblock {\em Commun. Math. Phys.} {\bf 265}, 1 
(2006).

\bibitem{PS2002a}
M.~Pr{\"a}hofer and H.~Spohn,
\newblock{\em In and out of equilibrium}, 
edited by V.~Sidoravicius (Birkh{\"a}user, Boston, 2002) 
p. 185. 

\bibitem{PS2004}
M.~Pr{\"a}hofer and H.~Spohn,
\newblock {\em J. Stat. Phys.} {\bf 115},  255 
(2004).

\bibitem{BFP2010}
{J. Baik, P. L. Ferrari and S. P\'ech\'e},
\newblock {\em Commun. Pure Appl. Math.}  {\bf 63},  1017 
(2010).


\bibitem{Miettinen2005}
L. Miettinen, M. Myllys, J. Merikoski and J. Timonen,
\newblock{\em Eur. Phys. J. B}  {\bf 46}, 55 (2005). 


\end{thebibliography}

%
%
%
%
\end{document}